\documentclass{WileyMSP-template}

\begin{document}

\pagestyle{fancy}
\rhead{\includegraphics[width=2.5cm]{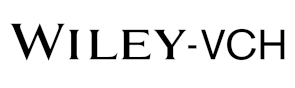}}

\title{Wafer-scale Graphene Electro-absorption Modulators Fabricated in a 300mm CMOS Platform}

\maketitle


\author{Chenghan Wu*}
\author{Steven Brems}
\author{Didit Yudistira}
\author{Daire Cott}
\author{Alexey Milenin}
\author{Kevin Vandersmissen}
\author{Arantxa Maestre}
\author{Alba Centeno}
\author{Amaia Zurutuza}
\author{Joris Van Campenhout}
\author{Cedric Huyghebaert}
\author{Dries Van Thourhout}
\author{Marianna Pantouvaki}



\begin{affiliations}
C. Wu,  Prof. D. Van Thourhout\\
Photonics Research Group, Department of Information Technology, Ghent University-imec, Technologiepark-Zwijnaarde 15, 9052 Gent, Belgium\\

C. Wu, Dr. S. Brems, Dr. D. Yudistira, Dr. D. Cott, Dr. A. Milenin, Dr. K. Vandersmissen, Dr. J. Van Campenhout, Dr. C. Huyghebaert, Prof. D. Van Thourhout, Dr. M. Pantouvaki\\
Imec, Kapeldreef 75, 3001 Leuven, Belgium\\

Dr. A. Maestre, Dr. A. Centeno, Dr. A. Zurutuza\\
Graphenea Semiconductor SLU, San Sebastian, Spain\\

Email Address: chenghan.wu@ugent.be; dries.vanthourhout@ugent.be; joris.vancampenhout@imec.be

\end{affiliations}


\keywords{CMOS-compatible, Graphene, photonics, integration, electro-absorption modulator}

\begin{abstract}

Graphene-based devices have shown great promise for several applications. For graphene devices to be used in real-world systems, it is necessary to demonstrate competitive device performance, repeatability of results, reliability, and a path to large-scale manufacturing with high yield at low cost. Here, we select single-layer graphene electro-absorption modulators as test vehicle and establish their wafer-scale integration in a 300mm pilot CMOS foundry environment. A hardmask is used to shape graphene, while tungsten-based contacts are fabricated using the damascene approach to enable CMOS-compatible fabrication. By analyzing data from hundreds of devices per wafer, the impact of specific processing steps on the performance could be identified and optimized. After optimization, modulation depth of 50 ± 4 dB/mm is demonstrated on 400 devices measured using 6 V peak-to-peak voltage. The electro-optical bandwidth is up to 15.1 ± 1.8 GHz for 25µm-long devices. The results achieved are comparable to lab-based record-setting graphene devices of similar design and CVD graphene quality. By demonstrating the reproducibility of the results across hundreds of devices, this work resolves the bottleneck of graphene wafer-scale integration. Furthermore, CMOS-compatible processing enables co-integration of graphene-based devices with other photonics and electronics building blocks on the same chip, and for high-volume low-cost manufacturing.
\end{abstract}


\section{Introduction}
\begin{figure}
  \includegraphics[width=\linewidth]{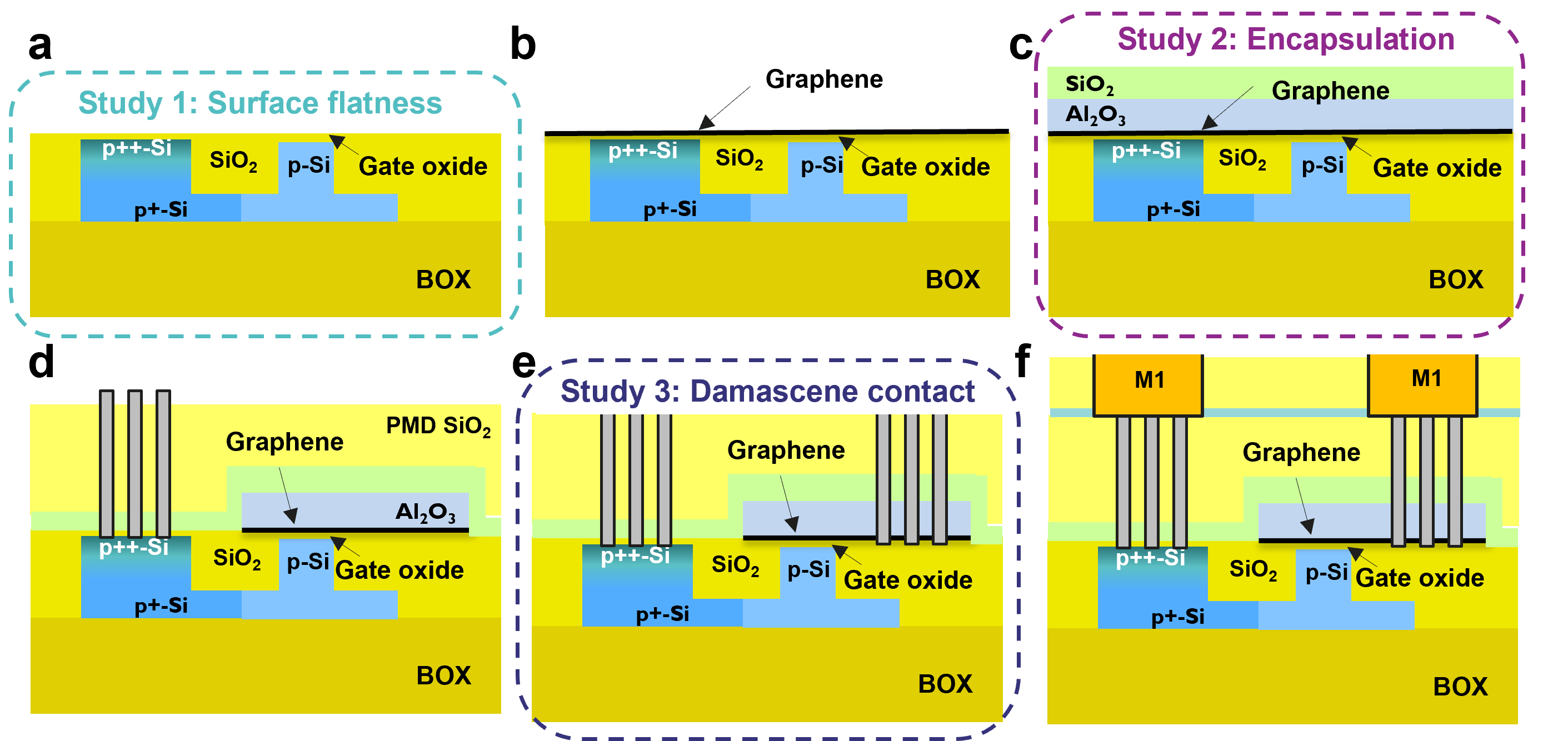}
  \caption{Proposed integration flow. (a) Waveguide patterning, surface planarization, and Si implantation steps, (b) wafer-scale graphene transfer, (c) graphene encapsulation, (d) graphene patterning and damascene contacts to p++ Si (e) graphene damascene contacts (f) final Cu metal lines.}
  \label{fig:fig1}
\end{figure}

Given its exceptional electrical and photonic properties\textsuperscript{\cite{neto2009electronic,mak2008measurement, xia2013interaction,bonaccorso2010graphene}}, graphene has attracted considerable attention in recent years. Its unique band structure and resulting broadband absorption spectrum, spanning from the ultraviolet to far-infrared\textsuperscript{\cite{nair2008fine}}, make graphene particularly promising for optoelectronic applications, while its large mobility\textsuperscript{\cite{banszerus2015ultrahigh,zomer2011transfer}} can be exploited in high-speed communications\textsuperscript{\cite{romagnoli2018graphene}}. However, its atomic layer thickness also limits the interaction strength with light. This constraint can be overcome by integrating graphene with Silicon Photonics: integrating graphene on a sub-micron scale waveguide and leveraging the evanescent field coupling, the interaction between the 2D-material and the light travelling through the waveguide can be enhanced. With this approach, integrated optoelectronic devices exhibiting ultrafast response and outstanding performance have been reported in recent years.\textsuperscript{\cite{agarwal20212d,goossens2017broadband,phare2015graphene,schuler2021high, Giambra2019}}\\
Silicon and silicon nitride based photonic integrated circuits are now being considered as a core technology for future optical interconnects, high-performance computing, light detection and ranging (Lidar) and sensing.\textsuperscript{\cite{reed2010silicon,thomson2016roadmap}} They can be fabricated at low cost and in large-volume with high-yield utilizing existing infrastructure of the complementary metal-oxide-semiconductor (CMOS) industry.\textsuperscript{\cite{rahim2018open,absil2015imec}} Graphene provides a number of advantages in terms of CMOS compatibility. Firstly, graphene itself is a CMOS-compatible material that can be grown by chemical vapor deposition (CVD) using wafer-scale tools\textsuperscript{\cite{verguts2016epitaxial,ramon2011cmos}}. Numerous research studies have been conducted on the large-scale growth of high-quality graphene.\textsuperscript{\cite{verguts2018growth,zhou2013chemical,wang2021single}} It has also been shown that graphene can be integrated by transfer onto almost any substrate as long as the surface is sufficiently flat, either in a single step using wafer-size graphene layers\textsuperscript{\cite{qing2020towards,lee2010wafer}} or in multiple steps using smaller patches to cover the entire wafer\textsuperscript{\cite{giambra2021wafer}}. This enables the integration of high-quality graphene on a silicon photonics platform in a straightforward manner. Moreover, while early demonstrations typically used doped silicon waveguides for controlling graphene’s electronic properties, more recent configurations consist of two graphene layers separated by a dielectric gate oxide that can be implemented on any type of waveguide, such as for example Silicon Nitride waveguides, therefore greatly enhancing its flexibility and eliminating the need for Si ion implantation. Finally, graphene transfer has a low thermal budget and can be performed in both the front- or back-end-of-line (FEOL or BEOL) in a CMOS process flow, which is advantageous for the cointegration with other silicon photonics modules.  In earlier work both waveguide integrated graphene modulators\textsuperscript{\cite{liu2011graphene,liu2012double,hu2016broadband}} and photodetectors\textsuperscript{\cite{pospischil2013cmos,Gan2013,Wang2013}} were shown.  Most of these demonstrations used small coupons or a non-scalable graphene supply. Recently, some promising results starting from 6” wafers\textsuperscript{\cite{giambra2021wafer,Schall2017}} and using scalable CVD-grown graphene were reported. Through systematic inline metrology, the quality of graphene was monitored at each stage of the process, at wafer-scale,\textsuperscript{\cite{giambra2021wafer}} bringing graphene-based photonics close to an industrial viable platform. However, none of this work was carried out using fully CMOS-compatible integration technology. \\
The primary issues are the lithography process, the graphene encapsulation and the graphene contacts.\textsuperscript{\cite{Mieikis2021}} At the moment, electron-beam (ebeam) lithography and lift-off-based contact metallization are mostly used but not compatible with high-volume industrial manufacturing.\textsuperscript{\cite{Mieikis2021}}  Standard photolithography utilizing a mask is preferred to enable high throughput and to keep the process cost effective, while typically a damascene process, involving via-etching, metal filling and planarization, is used for realizing contacts in CMOS fabs. To stabilize and protect graphene during further processing, a effective capping layer is another critical step that needs to be established using CMOS infrastructure. Therefore, the development of new and robust modules, adhering to the strict contamination requirements of CMOS fabs, are required for the scalable wafer-level integration of graphene based optoelectronic devices.\\
In this paper, we develop a wafer-scale integration process for realizing graphene-based photonics devices in a 300 mm CMOS pilot line.  As a test vehicle we choose an electro-absorption modulator (EAM) consisting of a doped silicon waveguide with a single layer of graphene integrated on top of a gate oxide, resulting in a graphene-oxide-semiconductor configuration.  The basic process flow\textsuperscript{\cite{wu2021graphene}} is outlined in Figure~\ref{fig:fig1}.  It consists of defining doped waveguides, including their planarization (Figure~\ref{fig:fig1}a), graphene transfer (Figure~\ref{fig:fig1}b), graphene encapsulation (Figure~\ref{fig:fig1}c), patterning of graphene and defining contacts to the doped silicon (Figure~\ref{fig:fig1}d), the graphene contacts (Figure~\ref{fig:fig1}e) and a final Copper damascene metal routing layer (Figure~\ref{fig:fig1}f).  In particular, we study and optimize three critical steps in this overall flow: the planarization step before graphene transfer (study 1), encapsulation of the graphene layer (study 2) and contacting the graphene layer using a damascene process (study 3).  Following optimization of these steps, hundreds of devices demonstrate performance comparable to that of lab-based devices with similar design and graphene quality\textsuperscript{\cite{alessandri2020high}}. The reproducible and robust integration route developed in this paper lays the groundwork for scaling also other graphene-based photonics devices and promoting their industrial adoption.

\section{Results  and  Discussion}

\subsection{Fab-level integration and optimization}
\begin{figure}
  \includegraphics[width=\linewidth]{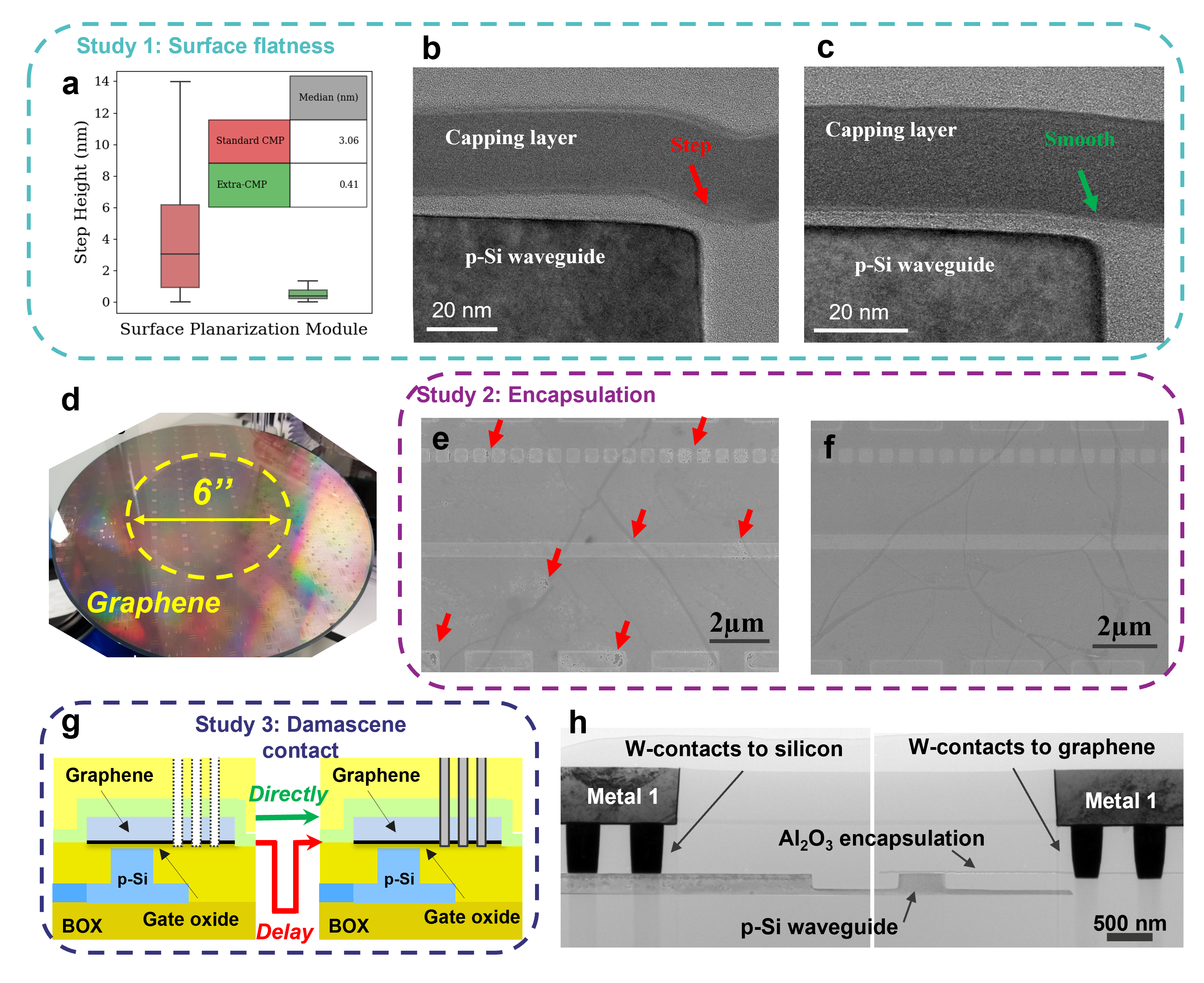}
  \caption{(a) Comparison of the step height remaining after surface planarization. Within 170 measured devices, the mean and standard deviation values of the step height are 4.3 ± 4.1 and 0.5 ± 0.3 nm for the standard CMP process and the process with the extra CMP step, respectively. Cross-TEM images taken at the waveguide edge for wafer with (b) standard CMP and (c) extra-CMP module. The standard planarization process results in a considerably higher remaining step and non-uniform gate oxide thickness. (d) Top-down image of 300 mm wafer with 6-inch graphene transferred at the center. Impact of soaking time. Representative top-down SEM image of wafer with (e) short and (f) long soaking time. Red arrows indicate the voids on top of the surface. The wrinkles in the graphene layer also visible in the pictures are induced during graphene growth and transfer. (g) Cross-section device scheme and description of the study on graphene contact. (h) Cross-section TEM of final device.}
  \label{fig:fig2}
\end{figure}

The integration flow started from 300 mm silicon-on-insulator (SOI) wafers with a 220 nm crystalline silicon layer and a 2 µm buried oxide (BOX). Standard 193 nm immersion lithography was used for pattering the silicon waveguides with a nominal width of 500 nm. One side of the waveguide was only partially etched to create a rib structure, allowing for electrical contacting through a 70 nm silicon slab layer. Afterwards, we utilized a standard chemical mechanical polishing (CMP) process, stopping on the SiN hardmask, a process also typically used in CMOS fabrication for shallow trench isolation (STI).\textsuperscript{\cite{seshan2012handbook}} Before removing the hardmask, we performed an oxide etch-back with diluted HF in an effort to lower the step height induced by the SiN mask removal. However, this approach typically results in a topography of a few nanometers locally at the edge of the waveguides. As graphene is a monolayer material, it is highly susceptible to its environment, and a few nanometers of step-height can already affect its properties and eventually device uniformity and yield across a 300 mm wafer. Therefore, we examined the impact of an extra CMP step designed to minimize the topography of the wafer prior to wafer-scale graphene transfer. After the hardmask removal, an additional oxide layer was deposited using a PECVD process on some wafers, followed by an extra CMP step stopping selectively on the Si waveguide. In Figure~\ref{fig:fig2}a, step height measurements at the silicon-oxide transition area show median values of 3.06 nm and 0.41 nm for the standard STI process and the process with the extra CMP step, respectively, confirming an improvement of surface flatness. The improved flatness of the wafer surface can also be observed from the cross-sectional transmission electron microscope (XTEM) images shown in Figure~\ref{fig:fig2}b and c. These images were taken after the full device fabrication. As indicated by the arrow in Figure~\ref{fig:fig2}c, the wafer with the additional CMP module has a more uniform and smooth oxide surface, especially near the waveguide edge. In contrast, the wafer with the conventional CMP module exhibits a discernible step at the side of the waveguide and a greater variation in the gate oxide thickness, which could result in larger strain in the graphene layer and a non-uniform electric field.  This will be elaborated further when discussing the results of Raman measurements and the electro-optical performance in the following sections.\\ 
Next, a 5 nm gate oxide was thermally grown on top of the waveguides. Three implantation steps were carried out, to minimize the contact and sheet resistance of the Si layers, without considerably increasing the optical loss in the waveguides. Then, a commercial company, Graphenea, grew a 6-inch graphene layer by chemical vapor deposition (CVD) and transferred it to the middle of a 300-mm wafer using a semi-dry technique as shown in Figure~\ref{fig:fig2}d. In this process, the graphene layer on its copper catalyst is attached to a polymer substrate, which allows to etch the copper catalyst away using a standard $FeCl_{3}$ wet etching method. After the etching, several consecutive ultra-pure DI water and acidic rinses were used to minimize Fe contamination. Graphene interface was then dried with N2 flow. When the graphene layer was dry, a dry lamination method was used to transfer the graphene onto the target wafers. The polymers/Graphene was laminated at a pressure above 1 bar and a temperature of 150\degree $C$ for the transfer. Finally, the remaining protective polymer layer is removed by a wet solvent process.\\
Given graphene’s self-passivated properties\textsuperscript{\cite{wang2008atomic,kim2014selective,karasulu2016continuous}}, it is difficult to directly deposit a dielectric on its surface. Commonly, a seeding layer is used to achieve homogeneous oxide deposition. Here we used a low-temperature surface physisorption based ‘soak’ method with tri-methylaluminium (TMA) as the precursor to carefully deposit a dielectric seeding layer. The actual $Al_{2}O_{3}$ capping layer was then deposited using a atomic layer deposition (ALD) process. To investigate the impact of the capping layer uniformity on the performance of the final devices, a second study was defined at this stage, whereby the soaking time was varied. Figure~\ref{fig:fig2}e shows a top-down scanning electron microscope (SEM) image after the PEALD deposition when short soaking time was used.  This image shows a large number of distinct voids in the $Al_{2}O_{3}$ layer. These voids could potentially lead to unintentional etching of the graphene layer during subsequent processing steps. Figure~\ref{fig:fig2}f, where a longer soaking time was applied, exhibits superior $Al_{2}O_{3}$ coverage of graphene, and the number of voids is significantly reduced. Only a few wrinkles generated during graphene growth and transfer remain visible. Overall, by optimizing the coverage of the capping layer, we expect to reduce the impact of later integration steps on the graphene layer achieving better device yield.\\
After deposition of the $Al_{2}O_{3}$ layer, a $SiO_{2}$ layer is deposited, also using a PEALD process, which is then patterned using DUV lithography and dry etching.  Following resist strip, the oxide layer is used as a hardmask to pattern the $Al_{2}O_{3}$ and graphene stack. Careful control of these steps is critical to avoid etching into the underlying silicon waveguides and is made possible through the use of high-end tools typical for a CMOS foundry. After graphene patterning, a pre-metal dielectric (PMD) is deposited and planarized by CMP, following a standard CMOS flow.\\ 
Finally, the contacts to both the graphene and the doped silicon layers are defined. The latter are fabricated first, by etching contact holes using reactive ion etching (RIE), which are then filled using a CMOS Ti/TiN/W damascene metallization process. A similar damascene process was used for contacting the graphene layer.  This is very different from most other work reported in literature, where typically a lift-off process is used to define top-contacts on graphene\textsuperscript{\cite{giambra2021wafer,alessandri2020high}}. Although this provides a low-cost and simple method for contact fabrication, it is not compatible with industrial CMOS process flows, where damascene processes are preferred as they offer higher yield and uniformity.  As selectively stopping the via etching process directly on top of the graphene layer would be very challenging, we choose to over-etch the oxide layer and create edge contacts.  Recent reports indicate that such an edge contact could offer lower contact resistance\textsuperscript{\cite{park2016extremely,Lee2022}}. The contact holes of 250nm diameter were patterned using DUV lithography and transferred in the PMD oxide by dry etching, selectively stopping on the $Al_{2}O_{3}$ capping layer. This step was then followed by resist striping, and etching of the $Al_{2}O_{3}$ and graphene layers, stopping in the underlying $SiO_{2}$ layer. Etching of graphene creates fresh dangling bonds, which can form strong covalent bonding with the metal\textsuperscript{\cite{robinson2010epitaxial,meersha2016record}} that is subsequently deposited. However, with increasing time elapse between the etching and metal deposition steps, these dangling bonds could bind with atmospheric water and oxygen and be passivated, hindering the formation of good contacts and increasing the resistance. The latter is detrimental for the high-speed response of the devices, as they are RC-limited. To study this effect in more detail, we kept this time-delay as short as possible for all wafers, except for one, for which we introduced an intentional gap of two days between the two steps, as illustrated in Figure~\ref{fig:fig2}g. Finally, the integration flow was completed with a conventional Cu-oxide metal-1 module. The final cross-section of the device is shown in Figure~\ref{fig:fig2}h. In this TEM image, the graphene layer is located below the $Al_{2}O_{3}$ capping layer. Notably, despite the fact that 6-inch graphene currently limits the number of available devices, the CMOS-compatible modules developed in this paper provide a 300 mm platform to scale up graphene-based photonics devices. Table~\ref{tbl:table1} summarizes the complete design of experiment (DoE) defined to study the effect of planarization, soaking time and contact module optimization.  The results from four wafers with this DoE, labelled wafers A, B, C, D, will be discussed in the following sections.\\

\begin{table}
  \caption{DOE summary of four wafers reported in this paper}
  \label{tbl:table1}
  \centering
  \begin{tabular}{ccccc}
    \hline
    DOE & Wafer A & Wafer B & Wafer C & Wafer D \\
    \hline
    Surface planarization & Standard STI & Standard STI & Extra CMP & Extra CMP   \\
    Encapsulation soaking & Short & Long & Long & Long  \\
    Contact metal deposition & No delay & No delay & 2 days delay & No delay  \\
    \hline
  \end{tabular}
\end{table}

\subsection{Raman characterization}
\begin{figure}
  \includegraphics[width=\linewidth]{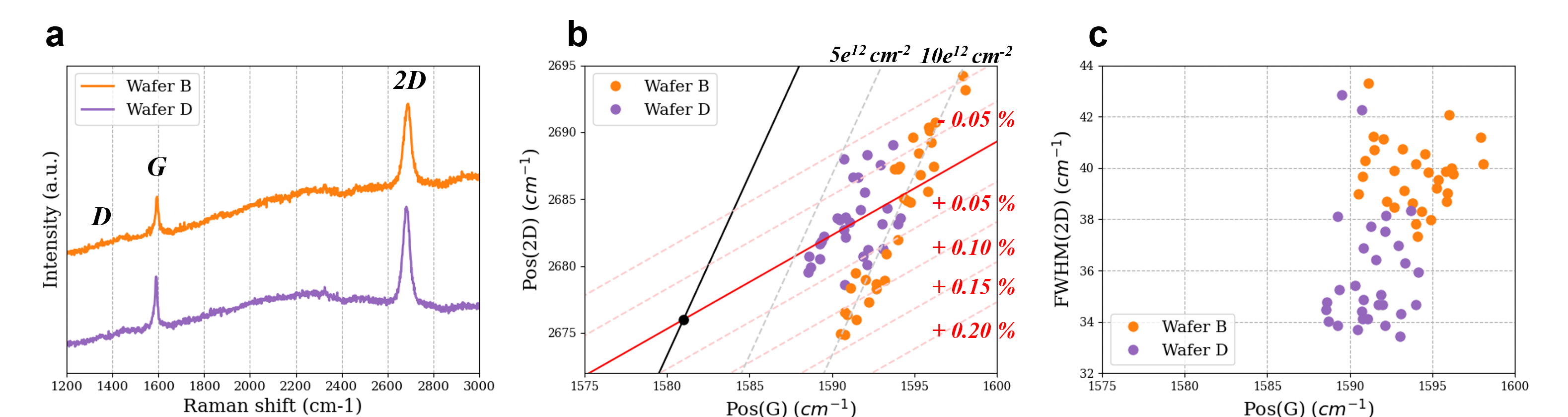}
  \caption{(a) Representative Raman Spectra for Wafers B and D after the full integration process. (b) Position and (c) FWHM of 2D peak as a function of the position of G peak. The black and red lines in Figure~\ref{fig:fig3}b are the theoretical trajectories indicating the effect of doping and biaxial strain, respectively. The back dot represents un-strained and un-doped graphene.}
  \label{fig:fig3}
\end{figure}

Before any electric-optical measurement, the graphene quality was checked by Raman Spectroscopy. Figure~\ref{fig:fig3} summarizes the most relevant results, focusing on the effect of the extra planarization step by comparing wafer B (standard CMP) and wafer D (extra CMP).  The measurements were carried out after completion of the full process flow, through the dielectric stacks of the metal-1 and PMD modules. From Figure~\ref{fig:fig3}a, the defect peak (D) is negligible for both wafers, confirming the proposed integration process does not result in significant degradation of the graphene quality. After fitting the G and 2D peaks of spectra taken at different locations within the wafer with a single Lorentzian, their relative position is mapped in Figure~\ref{fig:fig3}b. The black and red lines with slope of 2.745\textsuperscript{\cite{Das_2008}}, and 0.722\textsuperscript{\cite{Lee_2012}}, represent the effect of biaxial strain and doping respectively. The results indicate that the doping level of graphene varies from 6 to 10 × $10^{12}$ $cm^{-2}$ after the integration process. Wafer B suffers from more tensile strain effects (up to 0.14\%) compared to Wafer D (up to 0.07\%). Both wafers exhibit a similar amount of compressive strain, which could be explained by the deposition of the $Al_{2}O_{3}$ capping layer\textsuperscript{\cite{Robinson2011,Zheng2014}}. Figure~\ref{fig:fig3}c shows the full width at half-maximum (FWHM) of the 2D peak, with median values of 40 and 35 $cm^{-2}$ for Wafers B and D, respectively. These results verify that the smoother surface provided by introducing the extra CMP step is reducing strain effects and better preserves the quality of the graphene layer.\textsuperscript{\cite{Robinson_2009}}

\subsection{EO Static performance of inline EAMs}

\begin{figure}
  \includegraphics[width=\linewidth]{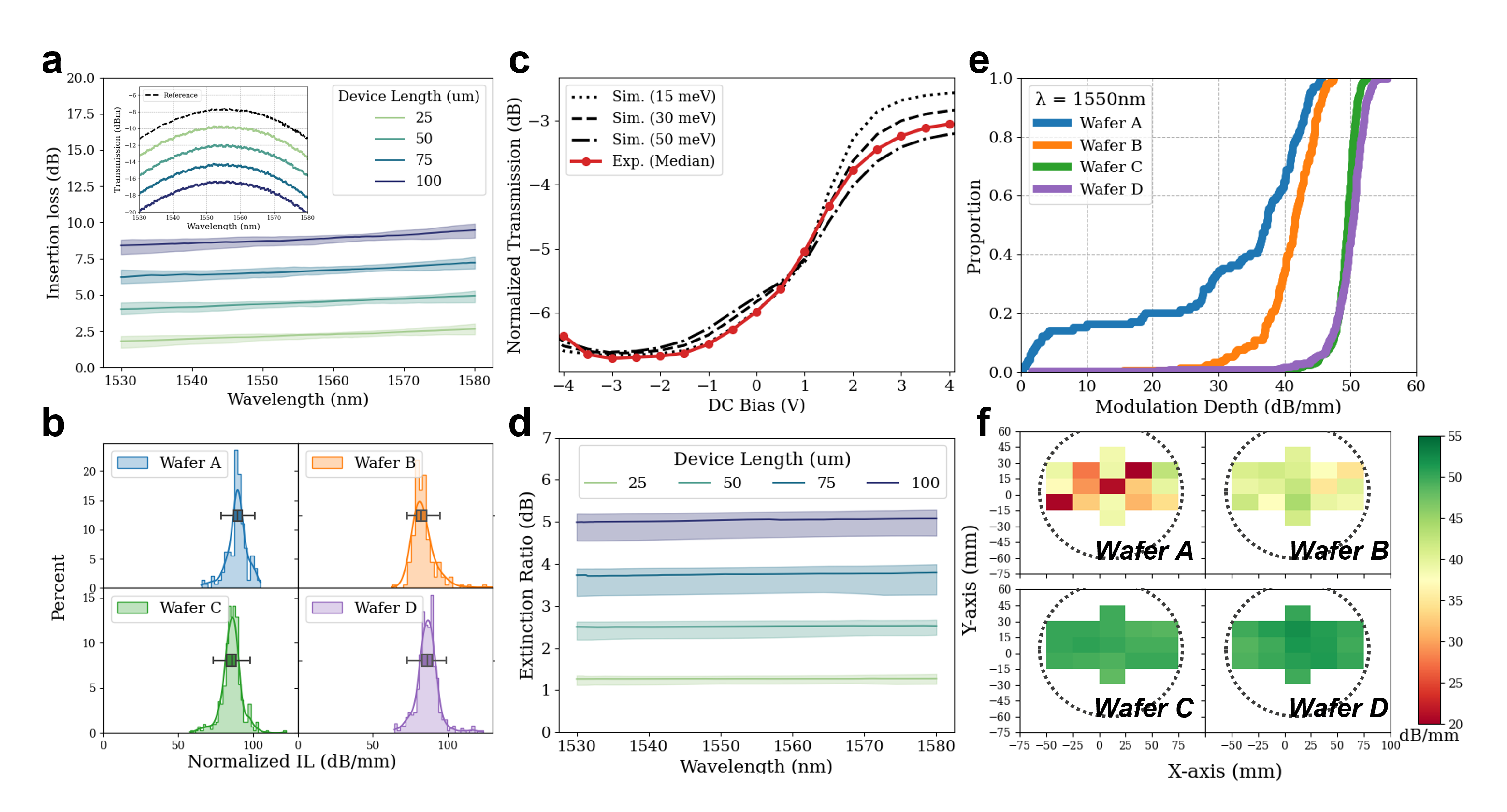}
  \caption{(a) Insertion loss as a function of wavelength for 25, 50, 75, and 100µm-long devices in Wafer D. The solid lines indicate the median value while the shaded areas show the 25-75 percentile. The inset shows representative transmission spectra of unbiased devices with different length. (b) Histogram of normalized insertion loss for all four wafers.(c) Normalized transmission of 75 µm-long devices of wafer D as a function of applied bias. Red solid line shows the median value of experimental results for 400 devices, black dashed lines represent simulation results for 3 different scattering rates. (d) Extinction ratio as a function of wavelength for 25, 50, 75, and 100µm-long devices (Wafer D). The solid lines represent the median value while the shaded areas show the 5-95 percentile of the results. (e) Cumulative distribution function and (f) wafer mapping of modulation depth at 1550nm wavelength for all four wafers.}
  \label{fig:fig4}
\end{figure}

The EAMs are designed for operation in the C-band with transverse electric (TE) polarization and coupled to an external laser source via grating couplers. In order to highlight the broadband nature of graphene, the wavelength was swept from 1530 nm to 1580 nm, for devices with four distinct device lengths. This wavelength range is restricted by the response of the grating couplers. The inset of Figure~\ref{fig:fig4}a depicts a representative transmission spectrum for all device lengths considered. By comparing with a neighboring straight waveguide without modulator, the loss from the grating coupler can be excluded and the wavelength dependent insertion loss (IL) can be determined. Figure~\ref{fig:fig4}a summarizes the IL for the unbiased devices measured across 17 dies of wafer D. The solid line represents the median value, while the band reflects the $25^{th}$ to $75^{th}$ percentiles for each active length. Next, we defined the normalized IL by comparing the peak transmission values of each curve and dividing by device length to capture the wafer-to-wafer variation in performance. Figure~\ref{fig:fig4}b shows a histogram of the normalized insertion loss for all 4 wafers. The mean values for wafers A, B, C, and D are 89 ± 7, 85 ± 12, 87 ± 7, and 87 ± 8 dB/mm, respectively. The comparable distribution in all four wafers suggests that graphene is transferred and patterned uniformly in each wafer, despite local variations in CVD graphene quality. Table~\ref{tbl:table2} provides a summary of the loss measurement data.\\
To evaluate the electro-optical (EO) response, a DC bias is then supplied to the devices.
Figure~\ref{fig:fig4}c shows a typical transmission response curve, measured at 1550 nm wavelength and normalized with respect to a straight waveguide. The red line represents the median value obtained from four hundred 75 µm-long devices measured on wafer D, whereas the black lines are simulation results generated by a commercial solver ($Lumerical^{TM}$) using three different graphene scattering rates. In the simulation, we set the doping level of the silicon waveguide and graphene layer at 1.5e18 $cm^{-3}$ and 1e13 $cm^{-2}$, respectively. We noticed that the curves generated by the simulation need a 1 dB downward shift in transmission and a -1.5 V shift in voltage to match well with the experimental results. The voltage adjustment can be explained by fixed charges inside the gate oxide, while the additional loss could originate from residues remaining after graphene transfer. The minimum transmission occurs at a negative voltage, indicating p-type doping of graphene.\\
Figure~\ref{fig:fig4}d shows the wavelength dependent extinction ratio (ER), for a 6V peak-to-peak drive voltage. The solid line and shaded range indicate the median and 5-95 percentile for each of Wafer D’s four different active lengths. We clearly observe that the EO response is consistently broadband and that the ER scales uniformly with device length, resulting in median values of 1.3, 2.5, 3.8, and 5.0 dB for 25, 50, 75, and 100 µm-long devices, respectively, at 1550 nm wavelength. To compare the DC performance between wafers, the modulation depth (MD), defined as the ER normalized by the active length, is calculated. The difference in performance and uniformity between the wafers is visualized by the cumulative distribution function (CDF) shown in Figure~\ref{fig:fig4}e. The mean and standard deviation values of the MD are 32 ± 13, 39 ± 4, 49 ± 2, and 50 ± 4 dB/mm for Wafers A, B, C, and D, respectively. The CDF curves in Figure~\ref{fig:fig4}e lead to three conclusions. (1) Despite the fact that the maximum MD of Wafer A and Wafer B are comparable, Wafer B has substantially lower variability. We ascribe this enhancement to the improved coverage of the capping layer, which minimizes the impact of following graphene integration processes. Overall, a longer soaking time and the resulting more uniform capping layer increased device yield by more than 20 percent and decreased the within-wafer standard deviation of MD. (2) Comparing wafer B (standard CMP) and wafers C and D (extra CMP) shows that the improved planarization boosts the modulation depth by 25\%. As indicated previously when discussing the Raman results, the smoother surface of wafers C and D reduces strain effects and better preserves graphene material quality, resulting in a larger ER within the same voltage range. In addition, the homogenous gate oxide can provide a constant electric field and uniform tuning of the graphene fermi level resulting in a steeper modulation response. (3) Finally, comparing wafers C and D, we can conclude that the DC performance is unaffected by the time delay introduced in the contact module, since both wafers exhibit a nearly identical CDF. Figure~\ref{fig:fig4}f depicts a wafer mapping of the modulation depth MD, with black dashed circles indicating the area where graphene was transferred. We measured devices on dies within a circular area with 75mm radius from the center of the wafer. Both wafers C and D exhibit excellent uniformity across 17 dies and 400 tested devices. On average, a modulation depth MD = 50 dB/mm is recorded, which is comparable to lab-based champion devices employing similar CVD graphene\textsuperscript{\cite{alessandri2020high}}. Wafers A and B on the other hand clearly exhibit less good uniformity and performance, which we attribute to the lower quality of graphene capping and planarization as discussed before. Table~\ref{tbl:table2} summarizes the results for extinction ratio and modulation response for all 4 wafers.

\begin{table}
  \caption{Summary of the static performance for all four wafers with four different active lengths.The unit of IL(ER) and normalized IL (modulation depth) are dB and dB/mm respectively.}
  \label{tbl:table2}
  \centering
  \begin{tabular}{ccccccc}
    \hline
    Wafer & IL-25$\mu{m}$ & IL-50$\mu{m}$ & IL-75$\mu{m}$ & IL-100$\mu{m}$ & Normalized IL & Observed devices \\
    \hline
    A & 2.2 ± 0.2 & 4.4 ± 0.3 & 6.6 ± 0.3 & 8.8 ± 0.2 & 89 ± 7 & 144   \\
    B & 2.4 ± 1.1 & 4.2 ± 0.5 & 6.3 ± 1.4 & 8.1 ± 0.5 & 85 ± 12 & 155  \\
    C & 2.0 ± 0.7 & 4.1 ± 0.5 & 6.3 ± 0.5 & 8.5 ± 0.5 & 87 ± 7 & 408  \\
    D & 2.1 ± 0.4 & 4.4 ± 0.7 & 6.8 ± 1.5 & 8.8 ± 1.7 & 87 ± 8 & 400  \\
    \hline
    Wafer & ER-25$\mu{m}$ & ER-50$\mu{m}$ & ER-75$\mu{m}$ & ER-100$\mu{m}$ & MD & Observed devices \\
    \hline
    A & 0.9 ± 0.3 & 1.6 ± 0.7 & 2.0 ± 1.1 & 2.9 ± 1.6 & 31 ± 14 & 100   \\
    B & 1.0 ± 0.2 & 2.0 ± 0.2 & 3.1 ± 0.3 & 4.1 ± 0.4 & 41 ± 5 & 155  \\
    C & 1.2 ± 0.1 & 2.5 ± 0.1 & 3.7 ± 0.2 & 4.9 ± 0.1 & 49 ± 2 & 408  \\
    D & 1.2 ± 0.2 & 2.5 ± 0.1 & 3.7 ± 0.3 & 5.0 ± 0.2 & 50 ± 4 & 400  \\
    \hline

  \end{tabular}

\end{table}

\subsection{EO Dynamic performance of inline EAMs}
\begin{figure}
  \includegraphics[width=\linewidth]{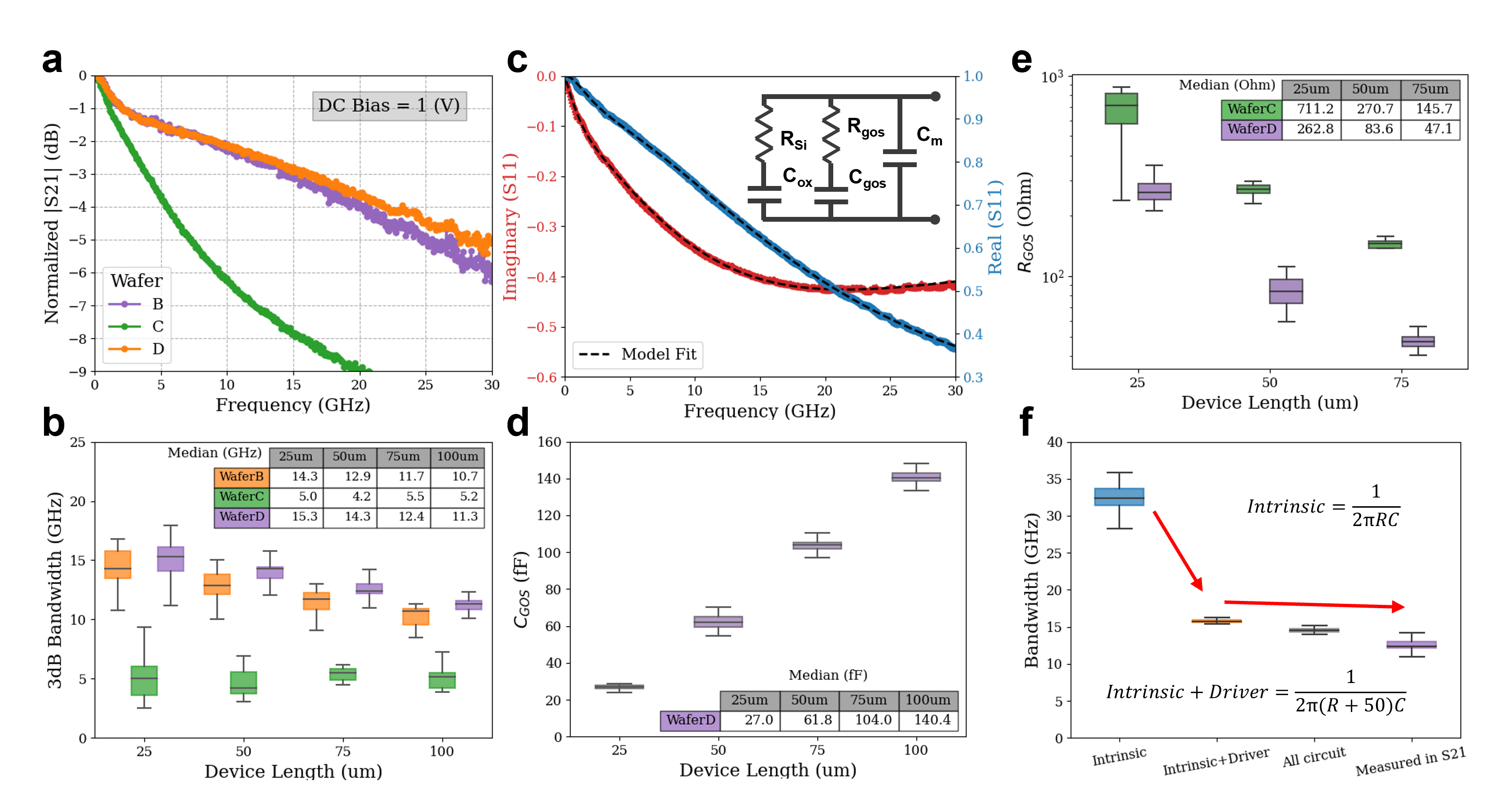}
  \caption{(a) Representative S21 response and (b) box plots of extracted EO bandwidth for wafers B, C and D at DC bias of 1V. Inserted table gives the median value for each device length and each wafer.(c) Representative S11 response and fitting results. The inset shows the equivalent circuit model of our structure, where $R_{Si}$, $C_{ox}$, $R_{gos}$, $C_{gos}$ and $C_{m}$ represent silicon resistance, oxide capacitance, GOS resistance, GOS capacitance and metal capacitance, respectively. (d) Box plots of extracted GOS capacitance $C_{gos}$ for wafer D. (e) Box plot of extracted GOS resistance $R_{gos}$ for 25, 50, and 75-µm long devices in Wafer C and D. $R^{2}$ values for the fit were larger than 0.9 and 0.98, respectively. The table in the inset shows the median values. (f) Bandwidth estimated from the fitting results, and bandwidth measured from S21 for Wafer D. The equations used to calculate these values are shown inside the figure.}
  \label{fig:fig5}
\end{figure}

we performed S-parameter measurements to assess the frequency response of the devices. An RF small-signal ranging from 100 MHz to 30 GHz was applied to the graphene modulators. A DC bias of 1 V is selected to ensure modulation at the slope of the transmission curve. Figure~\ref{fig:fig5}a shows a representative result for a 25-µm long device of wafers B, C and D. The 3dB-bandwidth for the wafer C device is 3.8 GHz, evidently much lower than for the other two devices (15.3 and 16.1 GHz for wafer B and D respectively). Figure~\ref{fig:fig5}b shows the statistics for all devices measured.  These reveal that wafer C, for which a delay was introduced between the contact etch and metallization process, has consistently a lower EO bandwidth than the other two wafers, for all four device lengths. It suggests that the time-delay during fabrication hinders good bonding between metal and graphene resulting in a higher contact resistance. This will be discussed further in the next section. For wafer D, median values of 15.3, 14.3, 12.4, and 11.3 GHz are measured for 25, 50, 75, and 100-µm long devices respectively, comparable with lab-based hero devices with similar design and graphene quality. To understand this length dependence better and get more insight on these devices, we further analyzed the S11 response for wafer B and C devices.\\
Since the dynamic response of our graphene modulator is primarily limited by the electrical RC constant\textsuperscript{\cite{alessandri2020high}}, we continue our analysis by fitting the S11 response to the equivalent circuit model shown in the inset of Figure~\ref{fig:fig5}c. The graphene-oxide-silicon (GOS) structure can be considered as a lumped device with a capacitance $C_{gos}$. The total resistance $R_{gos}$ of the device includes both contact and sheet resistance of silicon and graphene. $R_{si}$, $C_{ox}$, and $C_{m}$ are parasitic components, representing resistance of the substrate, capacitance of the buried oxide layer and capacitance of the metal pad, respectively. Figure~\ref{fig:fig5}c shows the S11 response for a 25-µm long device of wafer D, along with the result of the fitting process. From these, the total resistance $R_{gos}$, and capacitance $C_{gos}$ can be determined. Figure~\ref{fig:fig5}d summarizes the extracted device capacitance for Wafer D, which served as the basis for this analysis. As anticipated, $C_{gos}$ scales linearly with device length, resulting in wafer median values of 27, 62, 104, and 140 fF for devices that are 25, 50, 75, and 100 µm long, respectively. When evaluating Wafer C, the range of the capacitance was given to match the results obtained from Wafer D. This allowed to have reasonable value and prevent unrealistic outcomes. Figure~\ref{fig:fig5}e shows a wafer median value for the resistance $R_{GOS}$ of 711, 271, 146 Ω for Wafer C and 263, 84, 47 Ω for Wafer D, for 25, 50, 75-µm long devices, respectively. The smaller resistance in Wafer D confirms that the limited time between oxide etch and metal deposition better preserves the graphene contact quality, resulting in larger EO bandwidth. Lastly, we recalculated the electrical bandwidth of the devices based on the fitting results. Wafer D’s intrinsic RC bandwidth, considering only $R_{gos}$ and $C_{gos}$, attains wafer median values of 22, 31, 32, and 30 GHz, respectively, for devices measuring 25, 50, 75, and 100 µm in length. However, when the 50 Ω load resistance from the vector network analyzer (VNA) is considered, the calculated values (BW) are reduced to 19, 19, 16 and 13 GHz. These values are close to the final calculation, which takes into account all other parasitic components ($C_{ox}$ and $R_{si}$). Figure~\ref{fig:fig5}f summarizes the calculation for the 75-µm long device, showing the intrinsic 3dB-bandwidth (1/2$\pi$ $R_{gos}C_{gos}$) extracted from S11-measurements, the effect of the 50 Ohm load resistance, the effect of the parasitics and finally the measured electro-optical 3dB bandwidth. Table~\ref{tbl:table3} provides information for the other lengths. In general, the final electrical BW derived from S11 data is close to our experimentally measured EO BW, demonstrating the accuracy of our equivalent circuit model.\\

Following the discussion above, the EO bandwidth of our SLG EAM devices is mainly limited by the RC constant. Reducing the capacitance and resistance of the devices is key towards realizing a high-speed EAM. In recent work, large-area single-crystal graphene with 7.3 × $10^{3}$ $cm^{-2}V_{-1}s_{-1}$ mobility\textsuperscript{\cite{wang2021single}} and extraordinarily low contact resistance (23 Ω at room temperature) using a Ti-graphene edge contact configuration\textsuperscript{\cite{park2016extremely}} has been demonstrated, which would allow for devices with lower sheet and contact resistance in the future. Capacitance reduction, on the other hand, is not as straightforward. Reducing the capacitor surface or increasing the equivalent oxide thickness (EOT) of the gate oxide will both result in a lower capacitance but lead to a trade-off between bandwidth, modulation efficiency and drive voltage. Modulation efficiency and speed should be balanced for modulators driven at CMOS-compatible voltages (below 2 V for conventional CMOS circuitry). A possible solution to this conundrum is to enhance the mode interaction with graphene. By improving the interaction of light with graphene using TM polarization\textsuperscript{\cite{liu2011graphene,hu2016broadband}} or by constructing a double-layer structure\textsuperscript{\cite{agarwal20212d, phare2015graphene, Giambra2019,liu2012double}}, graphene-based modulators can modulate light effectively even for shorter devices. Our high-yield wafer scale integration approach is ideal for systematic exploration of these potential device architectures. 

\begin{table}
  \caption{Summary of the outcomes from S-parameter for Wafer D with four different active lengths.}
  \label{tbl:table3}
  \centering

  \begin{tabular}{cccccc}
    \hline
    S-parameter outcomes & Unit & 25$\mu{m}$ & 50$\mu{m}$ & 75$\mu{m}$ & 100$\mu{m}$\\
    \hline
    Measured EO BW & GHz & 15.1 ± 1.8 &	14.1 ± 1.4 & 12.6 ± 0.9 &	11.2 ± 0.7   \\
    Fitting result: C$_{gos}$ & fF & 26.7 ± 1.5 &	62.1 ± 3.7 &	102.8 ± 4.7 &	139.2 ± 7.9  \\
    Fitting result: R$_{gos}$ & $\Omega$ & 280 ± 61 & 86 ± 18 &	49 ± 6 &	38 ± 4  \\
    Intrinsic BW & GHz & 22.0 ± 3.1 & 30.6 ± 4.0 & 32.2 ± 0.4 & 30.5 ± 2.6  \\
    Intrinsic + Driver  & GHz & 18.5 ± 2.2 & 19.0 ± 1.3 & 15.8 ± 0.3 & 13.1 ± 0.8  \\
    Final estimated BW  & GHz & 16.8 ± 2.0 & 17.2 ± 1.2 & 14.5 ± 0.4	& 12.2 ± 0.7  \\
    Observed devices  & {} & 29 & 29	& 32 & 28  \\
    \hline

  \end{tabular}

\end{table}

\section{Conclusions}
To summarize, we have demonstrated the integration of single layer graphene electro absorption modulators in a CMOS fabrication environment. Damascene contact and hardmask lithography were used to build the wafer-scale devices in accordance with industry standards. Three critical processing steps are also studied in this work to determine their effect on device performance. We discovered that the surface flatness has a significant impact on the graphene quality and electric field homogeneity, both of which affect the modulation depth of the final device. Following that, the uniform capping layer reduces the impact of later integration steps on the graphene layer, resulting in increased device yield. Finally the time delay involved in constructing the damascene contacts affects the contact resistance and the 3dB bandwidth of the EAMs. After optimizing these three critical processing steps and implementing a CMOS-compatible dedicated integration approach, the device yield exceeds 95\% with loss, extinction ratio, and 3dB bandwidth values comparable to CVD graphene devices previously demonstrated in the lab\textsuperscript{\cite{alessandri2020high}}. We anticipate that the knowledge presented in this study can be extended and applied to a sophisticated building block library of graphene-based optoelectronic devices, that includes modulators, photodetectors, and sensors. This work will underpin the industrial adoption of graphene-based photonics devices, paving the way for the next-generation datacom and telecommunications applications.\\


\medskip
\textbf{Acknowledgements} \par 
We acknowledge funding from EU Horizon 2020 research and innovation program under grant agreement no. 881603 (Graphene flagship Core3) and from imec’s industry affiliation R\&D program on Optical I/O. We thank H.C. Tsai and S. Sergeant for their assistance in Raman measurements.

\medskip
\textbf{Data availability } \par 
The data that support the findings of this study are available from the corresponding author upon reasonable request.
\medskip

%
\bibliographystyle{MSP}
\bibliography{my_bib}



\begin{figure}
\textbf{Table of Contents}\\
\medskip
  \includegraphics{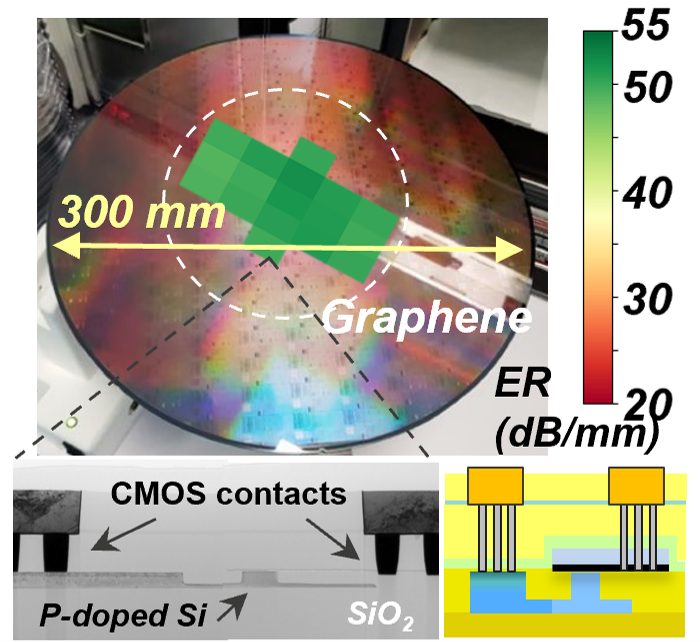}
  \medskip
  \caption*{This study describes the development of a CMOS-compatible integration for graphene-based photonics devices. The optimization of three critical processing steps has been explored. The results achieved are comparable to state-of-art devices. The reproducible and robust integration route developed in this paper lays the groundwork for scaling other graphene-based devices and promoting their industrial adoption.}
\end{figure}

\end{document}